\documentclass[lettersize,journal]{IEEEtran}
\usepackage{amsmath,amsfonts}
\usepackage{algorithmic}
\usepackage{algorithm}
\usepackage{array}
\usepackage[caption=false,font=normalsize,labelfont=sf,textfont=sf]{subfig}
\usepackage{textcomp}
\usepackage{stfloats}
\usepackage{url}
\usepackage{xcolor}
\usepackage{verbatim}
\usepackage{graphicx}
\usepackage{cite}
\usepackage{xcolor}
\usepackage{tabularx,booktabs,textcomp}
\usepackage{amsthm}

\usepackage{amsmath,amssymb,amsthm}

\begin{document}
\title{Detection of Malfunctioning Modules in Photovoltaic Power Plants using Unsupervised Feature Clustering Segmentation Algorithm}

\author{{Divyanshi Dwivedi, Pradeep Kumar Yemula,~\IEEEmembership{Member,~IEEE}, Mayukha Pal,~\IEEEmembership{Senior Member,~IEEE}}
 

\thanks{(Corresponding author: Mayukha Pal)}

\thanks{Mrs. Divyanshi Dwivedi is a Data Science Research Intern at ABB Ability Innovation Center, Hyderabad 500084, India and also a Research Scholar at Department of Electrical Engineering, Indian Institute of Technology, Hyderabad 502205, IN, (e-mail: divyanshi.dwivedi@in.abb.com).}

\thanks{Dr. Pradeep Kumar Yemula is an Assoc. Professor with the Department of Electrical Engineering, Indian Institute of Technology, Hyderabad 502205, IN, (e-mail: ypradeep@ee.iith.ac.in).}

\thanks{Dr. Mayukha Pal is a Global R\&D Leader – Data Science at ABB Ability
Innovation Center, Hyderabad-500084, IN, (e-mail: mayukha.pal@in.abb.com).}

}
\maketitle

\begin{abstract}

The energy transition towards photovoltaic solar energy has evolved to be a viable and sustainable source for the generation of electricity. It has effectively emerged as an alternative to the conventional mode of electricity generation for developing countries to meet their energy requirement. Thus, many solar power plants have been set up across the globe. However, in these large-scale or remote solar power plants, monitoring and maintenance persist as challenging tasks, mainly identifying faulty or malfunctioning cells in photovoltaic (PV) panels. In this paper, we use an unsupervised deep-learning image segmentation model for the detection of internal faults such as hot spots and snail trails in PV panels. Generally, training or ground truth labels are not available for large solar power plants, thus the proposed model is highly recommended as it does not require any prior learning or training. It extracts the features from the input image and segments out the faults in the image. Here we use infrared thermal images of the PV panel as input, passed to a convolutional neural network which assigns cluster labels to the pixels. Further, optimize the pixel labels, features and model parameters using backpropagation based on iterative stochastic gradient descent. Then, we compute similarity loss and spatial continuity loss to assign the same label to the pixel with similar features and spatial continuity to reduce noises in the image segmentation process. The effectiveness of the proposed approach was examined on an online available dataset for the recognition of snail trails and hot spot failures in monocrystalline solar panels.

\end{abstract}


\section{Introduction}
\label{section:Introduction}
\begin{figure}
\centering
  \includegraphics[width=3in]{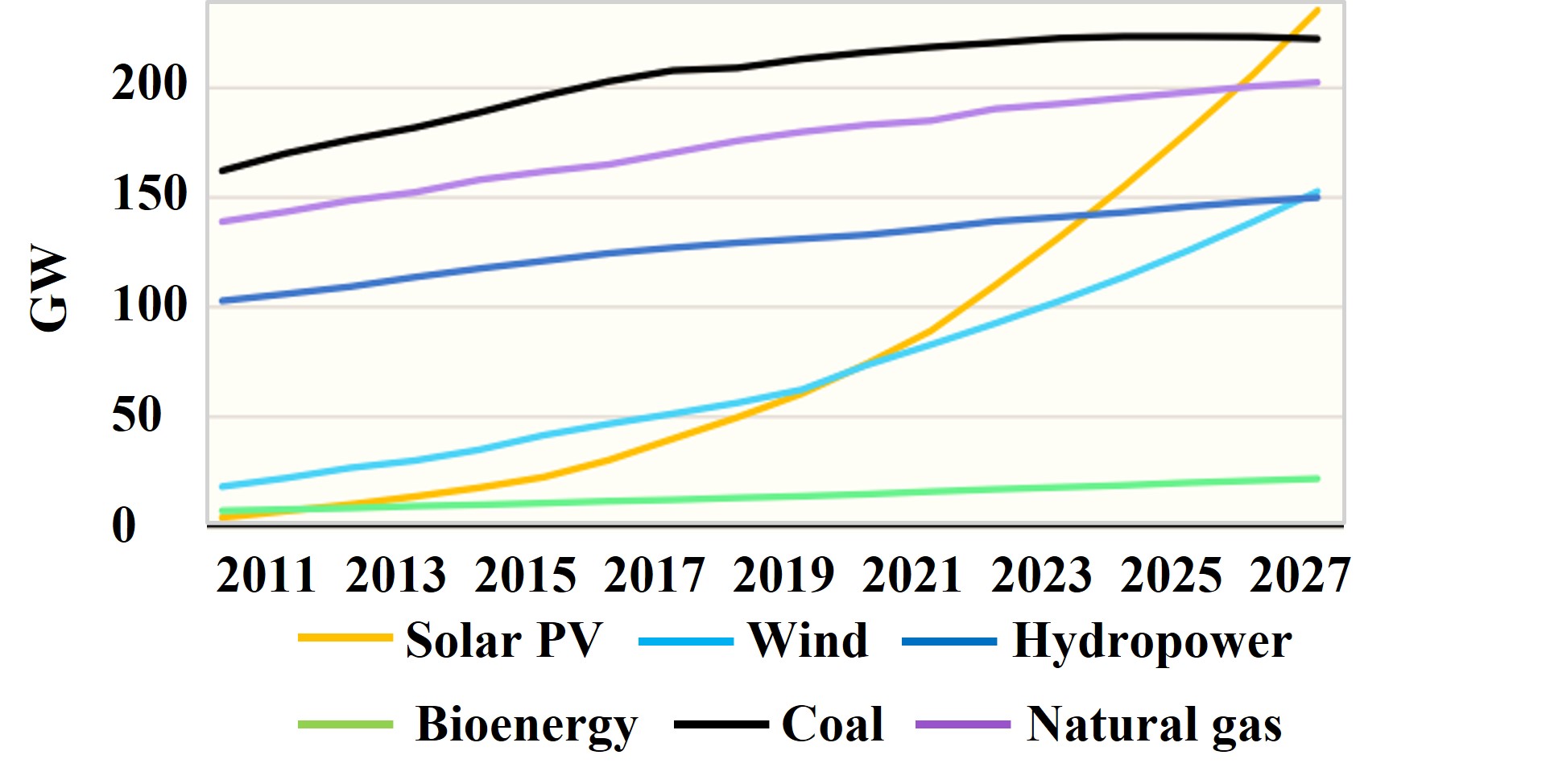}\\
  \caption{Cumulative power generating capacity of various sources from 2011- 2027}
  \label{fig:cumulative}
\end{figure}

\begin{figure}
\centering
  \includegraphics[width=2.8in]{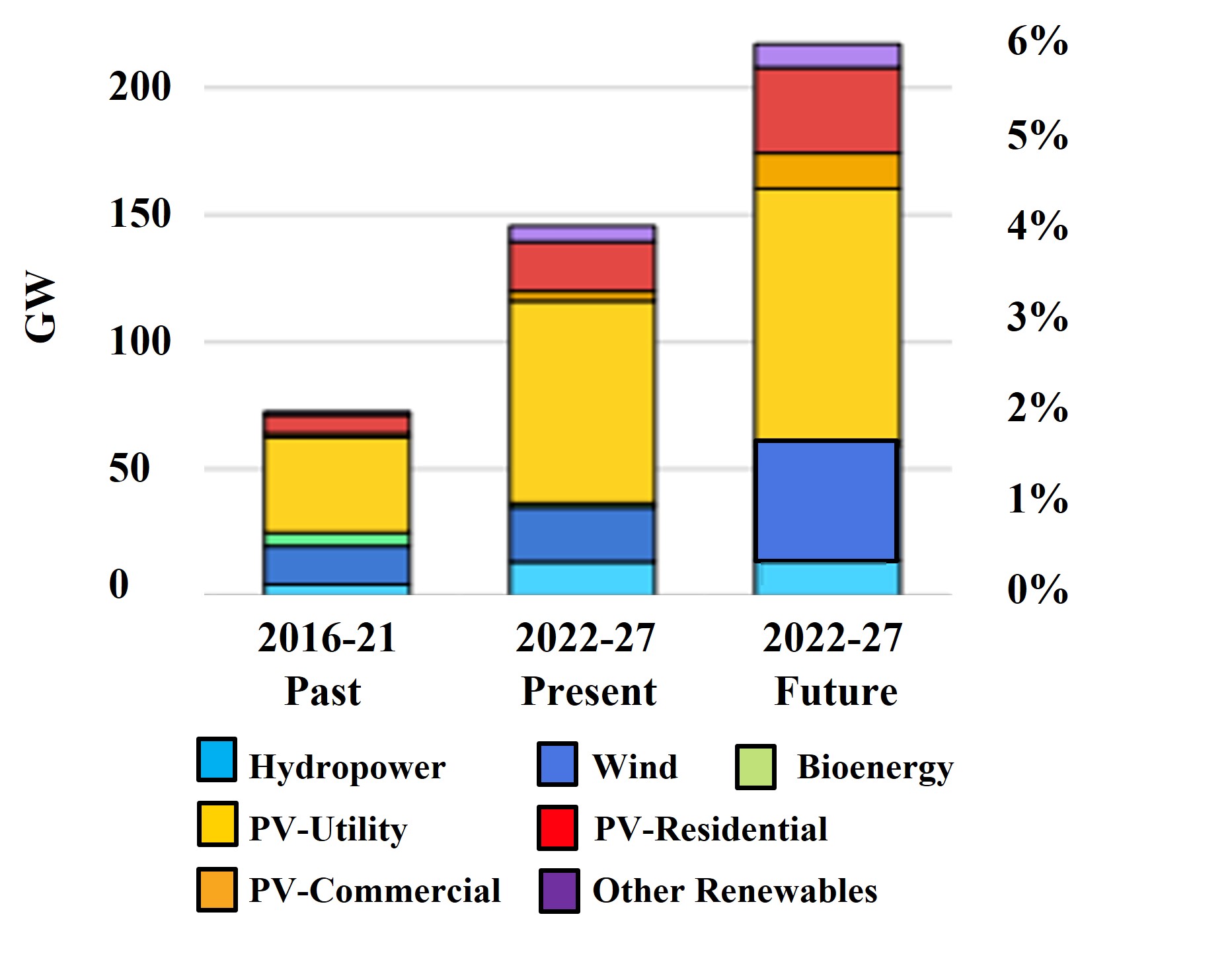}\\
  \caption{India’s renewable capacity additions from 2016-2027}
  \label{fig:India_stat}
\end{figure}

The increasing trend for more power generation utilizing renewable energy resources is born from the fact that conventional energy sources like coal, petroleum, and natural gas are on the verge of extinction \cite{segmentation_wiley}, \cite{GNN}.  Also, the global energy crisis provoked because Russia invaded Ukraine has given rise to the unrivalled momentum for renewables. Disruptions in supplying and high prices of fossil fuels during the crisis are the reasons that the countries are strengthening their policies for supporting more power generation using renewables. Experts are also predicting that throughout 2022-2027, renewables would account for over 90\% of global electricity expansion and surpass coal \cite{renewablestat}. The cumulative power generation using various renewable sources is shown in Figure \ref{fig:cumulative}, considering the past, present and future scenarios \cite{resiliency_cost}. For instance, India is set to double the new installations of solar photovoltaic panels to achieve the target of 500 GW by 2030 and its additions towards renewables are shown in Figure \ref{fig:India_stat}.

From Figure \ref{fig:India_stat}, we can say that renewable additions are majorly contributed by the solar PV generation installed from the utility end. The utility-scale solar photovoltaics is referred to as a large number of solar modules installed together to establish a power plant \cite{ViT}. Solar power plants require acres of land for encompassing. For example, Bhadla Solar Park, in Rajasthan, India is the world's largest solar power plant established in 2020 encompassing nearly 14,000 acres of land and generating 2.25 GW of power. However, after the commissioning of the huge power plants, their monitoring and maintenance become challenging. Monitoring is an important measure taken to increase the power output from solar PV panels, which can be affected by several factors such as shading, wear and tear due to environmental conditions \cite{resiliency}. To enhance the performance of solar PV panels and generate more power, it is feasible to implement smart monitoring which automatically detects the faulty cells in the solar PV panels without the requirement of large human interference as manual inspection of PV modules is simply not possible. Usually, these faults in solar PV panels are referred to as hot spots and potential induced degradation (PID). Hotspots occur when the solar panel is in shading condition and current flow is not possible across the weak cells and current concentration increases to other cells, leading it to overheat and mechanically damage \cite{segmentation_IEEE}. PID occur because of humidity, heat or voltage variations in the cell. As solar panels are designed with a life span of 25 years but these faults can sharply decrease their performance and efficiency. Thus, it is advisable to have a monitoring feature in large solar power plants to make its performance more reliable and durable.

The monitoring of solar PV panels is implemented manually using visual inspection or analysing through current-voltage characteristics \cite{IV_insepection} and another effective approach is to capture the infrared (IR) thermographic images of solar panels and identify the faults \cite{segmentation_springer}. The use of IR thermographic images is a reliable technique for the identification of faults in solar PV panels. Capturing the thermal images manually on the ground level is a time taking process as observed in \cite{8435954}. For 3 MW of solar power plants, analysis using ground images took 34 days of the inspection time; while capturing aerial thermal images reduced the time of analysis to approximately 3 hours. Thus, it is advantageous to prefer aerial thermal images for the analysis \cite{hotspot_define}. Various methods have been proposed to identify the faults in solar PV panels using aerial thermal images. For instance, the Canny edge image segmentation technique is used to identify the region of interest (ROI) i.e., affected region in solar panels \cite{canny}. The Robust Principal Component Analysis (RPCA) is used to separate the sparse corrupted anomalous components from a low-rank background \cite{pca}. Recent review works have been presented that detail the proposed methods and their inadequacy to address the problem \cite{review}, \cite{review2}, \cite{review3}, \cite{review4}. It has clearly stated the issues in the current state of the art because of the complexity and multi-layering process. 

Recently, many researchers have proposed frameworks using deep-learning algorithms where the infrared thermal imagery dataset is first labelled as faulty and healthy PV modules for autonomous PV module monitoring. Further, they train the model and using classification technique, faults have been identified. For instance, a binary classifier with a multiclass classifier is used to detect the fault and its type \cite{binary}. Convolutional neural networks and decision tree algorithms are used for the detection of external faults such as delamination, burn marks, glass breakage, discolouration, and snail trails on solar panels \cite{decision}. There has been a lot of work done in the same context \cite{ensemble}, \cite{HAIDARI2022102110}, \cite{cnn1}, \cite{NAVEENVENKATESH2022110786}, \cite{automatic}. All these works mentioned requires training labels or ground labels for training the proposed models; in a practical scenario, it is not always possible to have ground truth images of solar PV panels. Thus, these classification techniques are not adequate to rely on for real-time automatic monitoring of large-scale solar power plants. Another technique referred to as image segmentation is used to detect hot spots in solar panels. Otsu thresholding algorithm and its modified versions have been used to segment the faulty regions in solar panels but their segmentation accuracy is low, thus they are also not reliable \cite{otsu}, \cite{otsu1}. 

We propose an unsupervised method that does not require prior training labels or ground truth. The unmanned aerial infrared thermographic images of solar power modules in power plants are captured and directly fed as input. Then with the help of convolutional neural network layers, features of images are extracted and clustered to segment the objects based on the features of the images \cite{main}. This unsupervised learning optimizes the clusters by backpropagation using iterative stochastic gradient descent. Then with help of the loss function, we have also reduced the noises occurring during the segmentation process. Finally, we get segmented RGB images from which we can identify the hotspots, PID and snail trails in the solar PV modules. Further, these RGB images are converted to greyscale to enhance the features of the image so that faulty cells, normal cells and backgrounds are easily differentiable. Earlier, the proposed unsupervised feature clustering segmentation algorithm was found effective for bone age assessment to diagnose growth disorders using x-ray images \cite{xray}, to realize the extraction process of cage aquaculture \cite{aqua} and for segmentation of rock and coal images in the mining industry \cite{coal}.

The key contributions of this work are as follows:

\begin{itemize}
    \item A novel method is proposed for the identification of internal faults such as hotspots, snail trails and potential-induced degradation in solar PV panels.
    \item The input infrared thermal images are captured using unmanned aerial vehicles and fed to the unsupervised deep learning algorithm which performs segmentation based on feature clustering.
    \item It does not require any prior training labels and ground truth labels. Thus, the proposed algorithm is suitable to get integrated into any large-scale solar power plant.
\end{itemize}

The paper is organized as follows: a description of the dataset is provided in Section \ref{section:Datasets}, detailed explanation of the methodology of the unsupervised learning algorithm for segmentation is presented in Section \ref{section:Feature}. Section \ref{section:Result} provides the experimental result on a real dataset of solar PV panels. Finally, Section \ref{section:Conclusion} concludes the paper.

\section{Description of Dataset}
\label{section:Datasets}
Here, we used the online accessible dataset from \cite{Dataset}. This dataset comprises infrared thermal images of solar photovoltaic panels. The infrared thermographic surveillance technique is found to be effective for defect identification and analysis. Generally, objects with more than absolute zero temperature radiate thermal energy in infrared form; if temperature increases, radiated thermal energy gets more intense. Thus, these thermographic images allow the analysis of temperature variations. Thermal cameras are used for capturing infrared thermographic images \cite{hotspot1}. Thermal cameras can capture radiations in the infrared range of the electromagnetic spectrum. The zones with higher temperatures could be captured easily, but the human eye is not capable to identify these zones. Thus, image segmentation techniques are found suitable. For solar photovoltaic panels, infrared thermographic images can be captured using unmanned aerial vehicles (UAVs), which travel over photovoltaic panels by taking the proper angle setting of the camera into consideration.

\begin{figure}
\centering
  \includegraphics[width=2in]{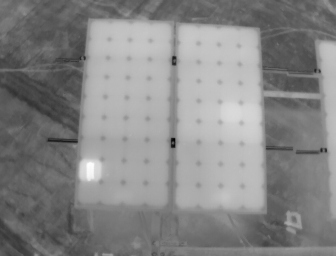}\\
  \caption{Grey-scale image pre-processed from a thermal image of solar PV panel}
  \label{fig:thermal}
\end{figure}

\begin{figure}
\centering
  \includegraphics[width=2.2in]{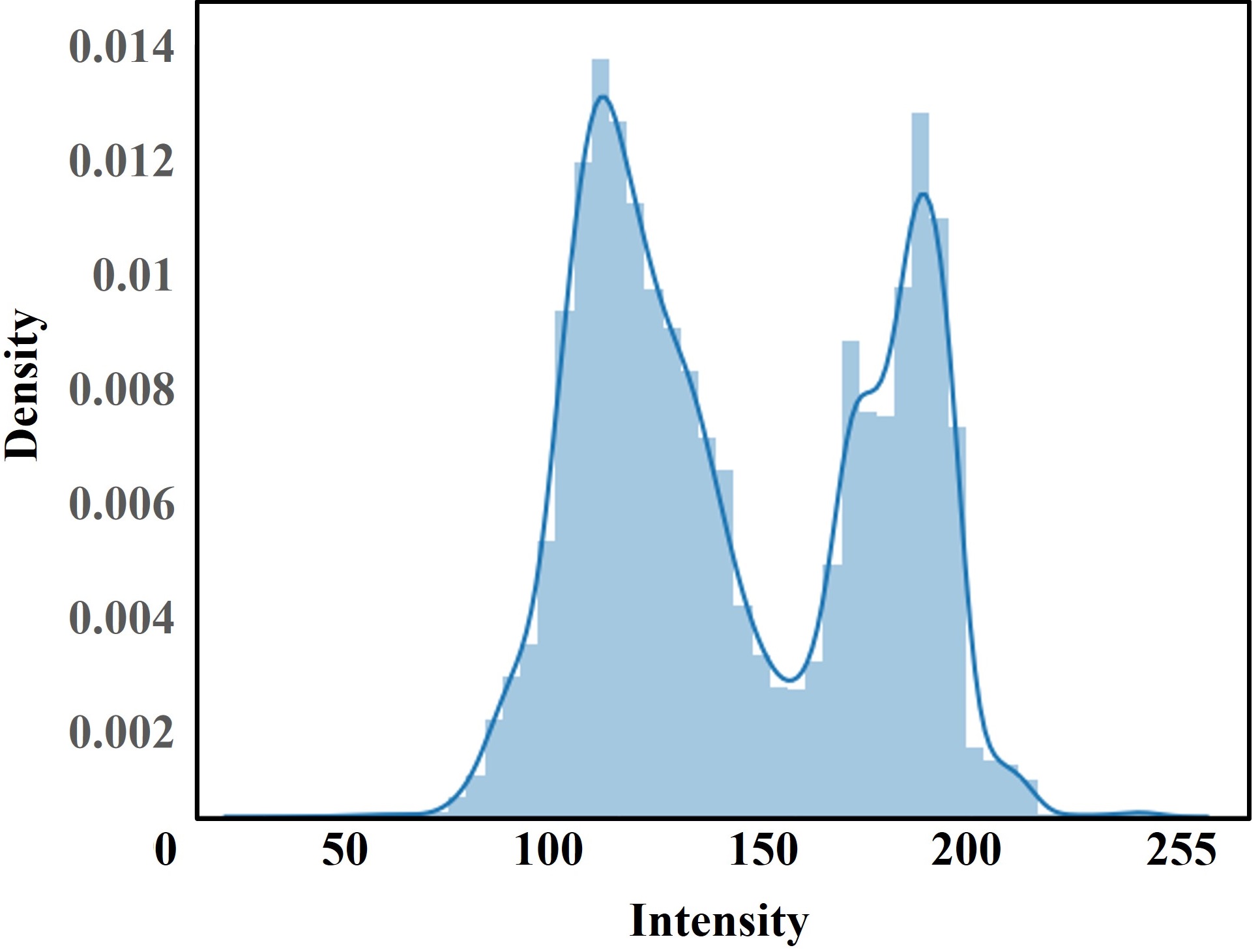}\\
  \caption{Histogram distribution of the grey-scale image of solar PV panel}
  \label{fig:histogram}
\end{figure}

The considered dataset was captured by Zenmuse-XT with the spectral range of 7.5 - 13 $\mu m$, thermal sensitivity $< 50\ mK$ and provided images of size $336 \times\ 256$ in JPG format. This camera glides over the monocrystalline solar panels and captures thermal images. Further for enhancing the features of the thermal images, they are pre-processed and converted into grey-scale images as shown in Figure \ref{fig:thermal}. Figure \ref{fig:histogram} shows the histogram distribution of the grey-scale image. It shows how many times each intensity value in the image occurs. The human eye is not capable to identify the fault arising in solar PV panels such as hot spots or snail trails present in the images of a photovoltaic panel. To solve this issue, we propose a deep learning-based segmentation algorithm to identify the hot spots in solar photovoltaic panels. These image datasets do not have any prior training labels available, thus unsupervised learning algorithm is best suited to identify the fault. In the real world also, the implication of the proposed model is feasible as it does not require any prior ground truth labels.

\section{Methods and Materials}
\label{section:Feature}

The IR thermal image dataset of Solar PV panels is of diversified resolutions, it is required to identify the hot-spot and snail trail regions while suppressing the noise. We exploit an unsupervised deep learning algorithm, inspired by a novel and effective unsupervised learning image-segmentation algorithm proposed in [1]. Implementation of supervised learning algorithms involves pixel levelling, ground truth images and original images. On the other hand, unsupervised algorithms require no prior labelling, training images or ground truth images; they extract the features of an image and implement the feature learning process to allocate the labels. Thus, we can easily identify the defects in the solar PV panels by segmenting images based on differentiable feature clustering. Here, we use Convolutional Neural Network (CNN) to extract the pixel-level features and identify the hot spots and snail trails in the input IR image of solar PV panels.

\begin{figure*}[]
\centering
  \includegraphics[width=6in]{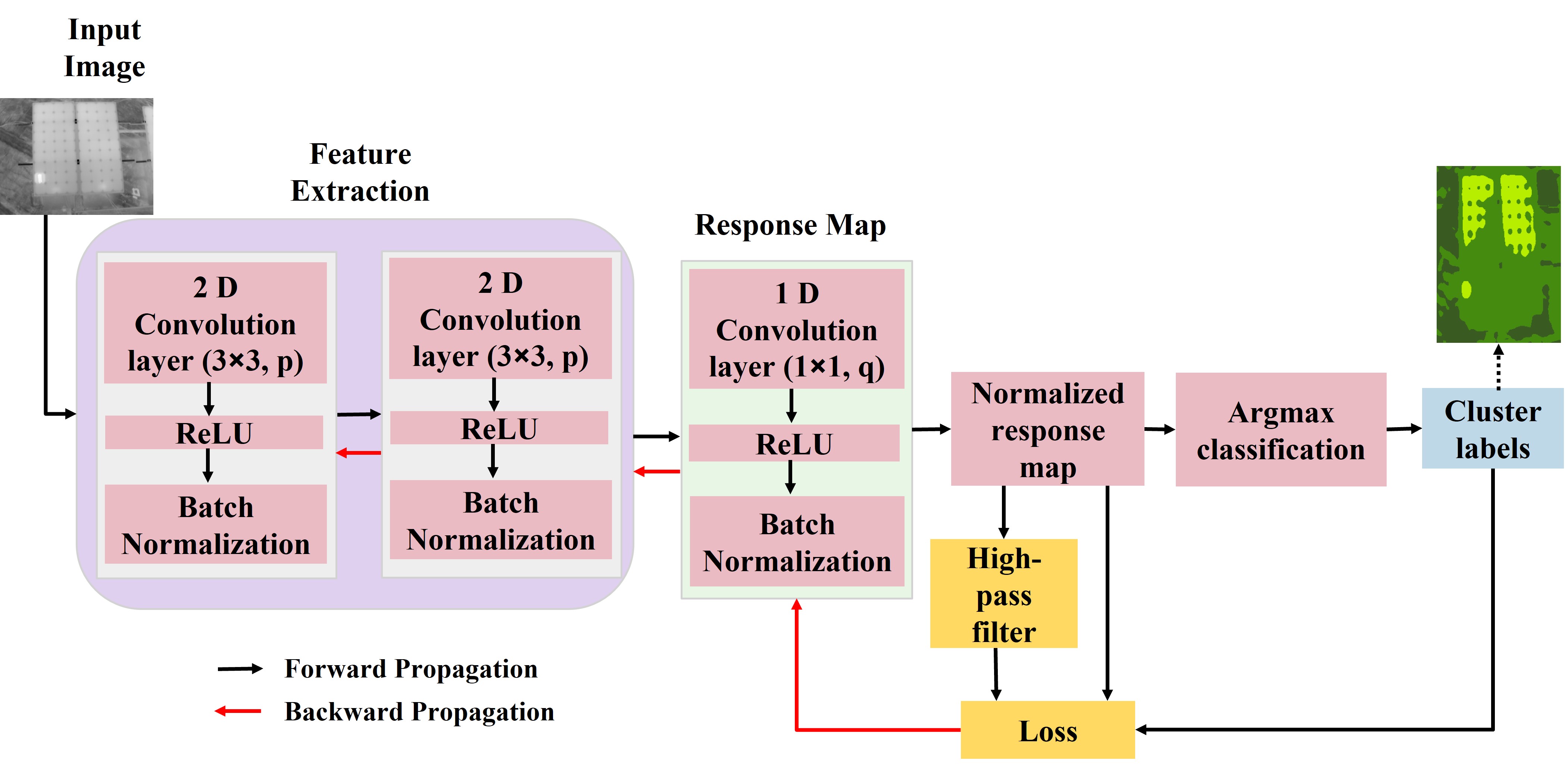}
  \caption{Architecture of the unsupervised image segmentation model for PV panels}
  \label{fig:Method}
\end{figure*}
The flow of the proposed algorithm is shown in Figure \ref{fig:Method}, includes CNN layers to fragment out the high-level features, a batch normalization layer to rescale and make the algorithm perform faster in stable mode, an argmax layer for pseudo labelling to the features extracted and backpropagation of loss function to evaluate the performance of the algorithm.

\subsection{Feature Extraction using CNN}

An image \{$I_{PV}\in \mathbb{R}^X$\} having dimension $L \times W$, where $k_n \in \mathbb{R}^X $ is the pixel value normalized to [0,1] is fed as input to the initial layer of the algorithm. A $p$-dimensional feature vector \{$m_n \in \mathbb{R}^p$\} is computed from $k_{n} $, by passing it through $M$ number of channels of two-dimensional convolutional layers of kernel size $3 \times\ 3$, RELU activation function, and followed by batch normalization layer to attain $N$ pixels of an input image. Subsequently, the feature vector \{$m_n$\} is fed as input to $q$ number of channels of the 1D convolutional kernel of size $1 \times\ 1$, then passed through the batch normalization layer. Finally, with a linear classifier, a response vector \{$rv_n \in W_l k_n $\} is obtained where $W_l\in \mathbb{R}^{q \times\ p}$, which is further normalized to $rv^\prime_{n}$ such that it has zero mean and unit variance \cite{main}. Then, the argmax layer is applied for labelling the clusters $c_n$ for each pixel by considering the criteria of selecting the dimension that has a maximum value of $rv^\prime_{n}$. Simply, we can say that the cluster label for the pixel was given which is equivalent to the maximum value of $rv^\prime_{n}$ and signifies the clustering of the feature vector into $q$ clusters. The $t-th$ cluster of the endmost response $rv^\prime_{n}$ is given as:

\begin{equation}
    C_t = max \{ rv^\prime_{n}\}=  {rv^\prime_{n}} \in \mathbb{R}^q  \hspace{0.25cm}| \hspace{0.25cm} rv^\prime_{n,s} \leq {rv^\prime_{n,t}}, \hspace{0.25cm}\forall s   
\end{equation}
where, ${rv_{n,t}^{'}}$ represents the $t-th$  element of ${rv_{n}^{'}}$. This process is equivalent to the assignment process of each pixel to the neighbouring nodes of the $q$-points which are present at infinite distances, in $q$-dimensional space on the relevant axis.

\subsection{Number of Clusters}

In unsupervised learning of image segmentation, the number of definite labels for the cluster varies on the image fed as input. If an image has a multitude of objects then it will have a high number of clusters and vice versa. Initially, the number of clusters $q$ for training the model is kept high. Then with the effective use of feature similarity and spatial continuity constraints, similar and neighbouring pixels are merged and eventually end up with a smaller number of clusters. Here, we considered 18 and 4 as the maximum and the minimum number of clusters respectively for segmenting the image into hot spots, snail trails, background, and normal regions of solar PV panels.

\subsection{Loss Function}

The loss function is used as a constraint for improving the feature similarity and spatial continuity between the pseudo labels assigned to image pixels, which is given as:
\begin{equation}
    L=L_{fs}+\alpha L_{sc}= L_{fs}\{{rv^\prime_{n}},c_n\}+ \alpha L_{sc}\{{rv^\prime_{n}}\}
\end{equation}

 where $L_{fs}$ is similarity loss and $L_{sc}$ is the continuity loss. $\alpha$ is weight balancing between the feature similarity and spatial continuity loss functions. As discussed in the previous section argmax function provides pseudo labels to the image’s pixels as per the features. After the assignment of pseudo labels, it is passed through this loss function.

\subsubsection{Feature Similarity Constraint}

Implementation of this constraint would help in stabilising the clusters in the image by enhancing the similarity of similar features. The image pixels that have similar features should be within a cluster and various clusters should have distinct features. The network weights are adjusted to minimize the similarity loss function to extract the important features for clustering. The feature similarity is computed using the cross-entropy loss between $rv^\prime_{n}$ and $c_n$ as:

\begin{equation}
  L_{fs} (rv^\prime_{n},c_n)= \sum_{n=1}^{N} \sum_{z=1}^{q} -\delta (z-c_n) \; ln(rv^\prime_{n})
\end{equation}

where,
$\delta(z)=\left\{
    \begin{aligned}
        & 1 \qquad  \textup{if} \quad z=1 \\
        & 0 \qquad  \textup{if} \quad z\neq 1 \\
    \end{aligned}\right.$

\subsubsection{Spatial Continuity Constraints}
It is preferred to have spatial continuity among the clusters of the image’s pixels as it helps to suppress the excess number of labels created due to the complicated structures and patterns in the image. Spatial Continuity Constraint is computed by taking the L1-norm of vertical and horizontal variation of response map $rv^\prime_{n} $  into consideration; implemented by a differential operator. Mathematically, it is defined as:
\begin{equation}
   L_{sc} ({rv^\prime_{n}})= \sum_{\beta=1}^{W-1} \sum_{\gamma=1}^{L-1} ||{rv_{\beta+1,\gamma}^{'}} - {rv_{\beta,\gamma}^{'}} ||_1 + ||{rv_{\beta,\gamma+1}^{'}} - {rv_{\beta,\gamma}^{'}} ||_1
\end{equation}

where, $L$ and $W$ are the length and width of an input image. ${rv_{(\beta,\gamma)}^{'}}$ is the pixel value at $(\beta,\gamma)$ from the response map $rv^\prime_{n}$.

\subsubsection{ Mechanism by Backpropagation}
This section details the approach for training the network in unsupervised image segmentation. After feeding the input image, with constant model parameters, cluster labels are predicted and then the model is trained based on the parameters by considering the predicted labels. The prediction of cluster labels is a forward process on the other hand training of the model is a backward process which is based on stochastic gradient descent with momentum by updating the model parameters. This stochastic gradient descent with momentum helps in accelerating the gradient vectors in the right direction, leading to faster convergence.  We are computing the loss and backpropagating it for updating the parameters. This forward and backward process is implemented in the loop for $E$ iterations to achieve the final segmentation of the image in clusters.\\

\begin{figure}
  \includegraphics[width=3in]{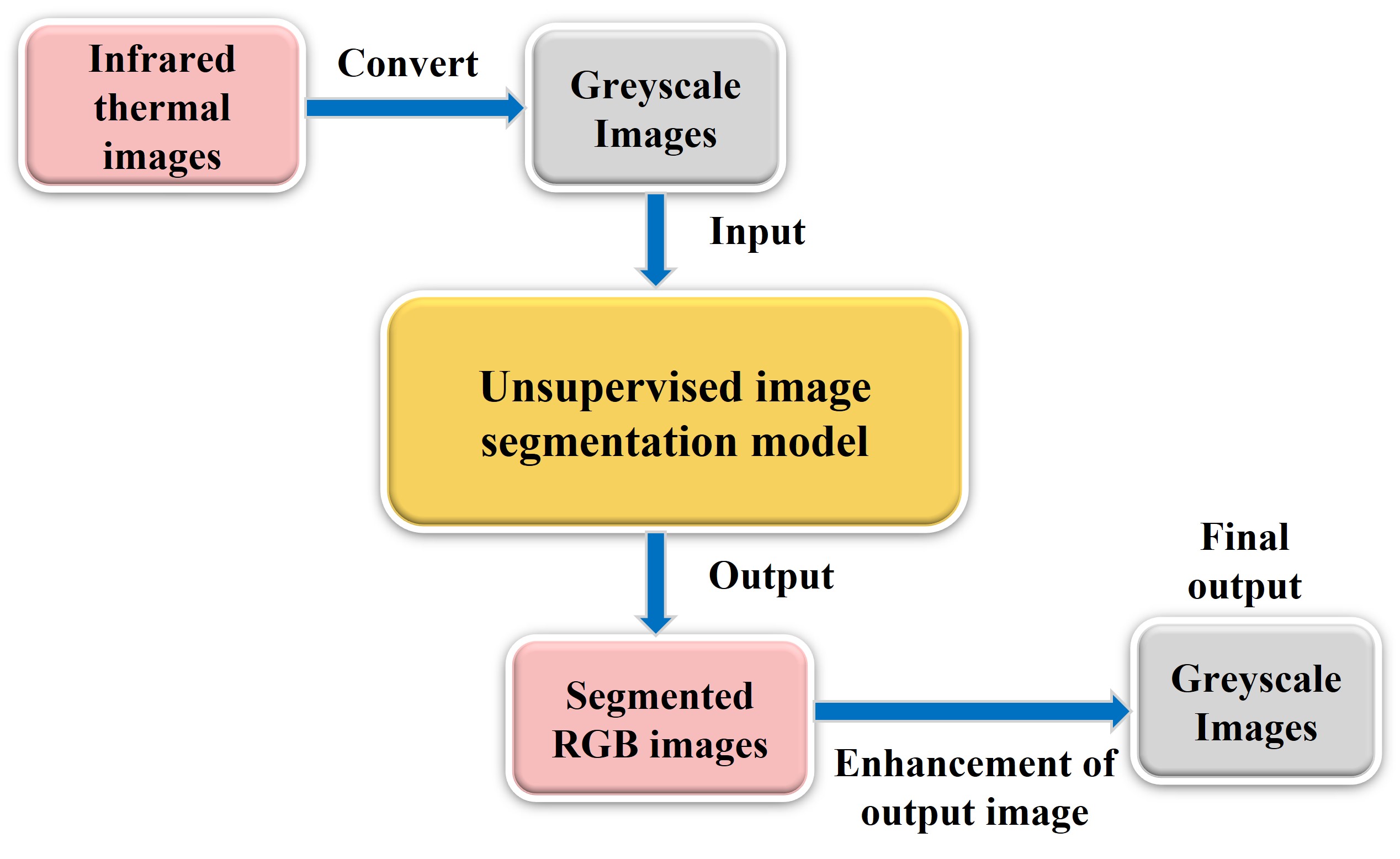}\\
  \caption{Process flow diagram for the proposed algorithm}
  \label{fig:Flow}
\end{figure}

The identification of similarity between features of various pixels is the first criterion that needs to be addressed for the allocation of labels to the pixels. As discussed above, first we fed the infrared thermal image of solar PV panels of size 336×256 to CNN modules for feature extraction. These CNN modules comprise a convolutional layer, a ReLU layer, and a BN layer and these layers are connected from end to end. We considered three CNN modules. The first two modules are equipped with two-dimensional convolutional layers of kernel size 3×3 and the last module has a one-dimensional convolutional layer of kernel size 1×1. Then it is passed through the argmax layer for pseudo labels. Further, the network is trained by computing the loss function and implementing the backpropagation to enhance the cluster segmentation in the image. All the parameters used for the segmentation of PV panel images are tabulated in Table \ref{tab:hyperparameters}. The chosen value for the weight balancing constant in the loss function is based on the loss variation with iterations and found the minimum value of loss when it is taken 5; discussed in detail in section \ref{section:Result}. Other parameters are chosen based on the standardized case for computational ease as in \cite{aqua}.
\begin{table}
  \centering
  \caption{Hyperparameters of unsupervised segmentation algorithm}
\begin{tabular}{cc}
 \toprule
Hyperparameters      & Values    \\
 \midrule
Size of IR thermal image of solar panel     & 336×256               \\
Stochastic gradient descent momentum       & 0.9               \\
Learning Rate  & 0.1               \\
Number of iterations       &  200               \\
Weight balancing constant in loss function  & 5\\
\bottomrule
\end{tabular}
\label{tab:hyperparameters}
\end{table}
\begin{figure}[h]
\centering
  \includegraphics[width=1.5in]{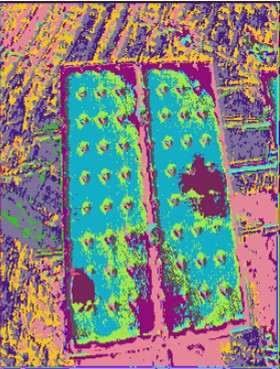}\\
  \caption{Initial pseudo cluster labels allotted to input thermal image of solar PV panel}
  \label{fig:forward}
\end{figure}

The image obtained by the implementation of the proposed algorithm would be in RGB colour. Thus, to identify and enhance the feature of RGB colour images, they are converted to greyscale images to detect the faults in the solar PV panels. The process flow of images is illustrated in Figure \ref{fig:Flow}.

\begin{algorithm}
\caption{Unsupervised segmentation of solar PV panels}\label{alg:alg1}
\begin{algorithmic}
\STATE 
\STATE {\textsc{Input:}} ${I_{PV}\in \mathbb{R}^X}$ with dimension L×W
\STATE {\textsc{Output:}} Hot spots in solar PV panels
\STATE {\textsc{Initialize:}} $E \leftarrow $ The number of iterations
\STATE \hspace{0.5cm}  Feature Vector $\{m_n\} \leftarrow$ 2D convolutional $\{k_n\}$
\STATE \hspace{0.5cm}  Response vector $\{rv_n\} \leftarrow$ 1D convolutional $\{m_n\}$
\STATE \hspace{0.5cm}  $\{rv^\prime_{n}\} \leftarrow Norm \{rv_n\}$
\STATE \hspace{0.5cm}  $\{c_n\} \leftarrow Argmax \{{rv_{n,t}^{'}}\}$
\STATE \hspace{0.5cm}  Compute $L$ using equation (2)
\STATE \hspace{0.5cm}  2D convolutional layers,\\ \hspace{0.5cm} 1D convolutional layer $\leftarrow$ Update ${L}$
\STATE \hspace{0.5cm}  Segmented RGB Image
\STATE \textsc{Return:}  Segmented Greyscale Image
\end{algorithmic}
\label{alg1}
\end{algorithm}

\section{Results and Discussion}
\label{section:Result}
In this section, we will discuss how the proposed model works for datasets detailed in section \ref{section:Datasets}. The available images of solar PV panels in the dataset are converted form of infrared thermal images into greyscale images. These greyscale images are considered as the input image for the proposed model. For analysis, we used six images from the dataset and fed them independently to the proposed unsupervised image segmentation algorithm. In detail, we have already discussed the methodology of the proposed algorithm in section \ref{section:Feature}.

\begin{figure}[h]
\centering
  \includegraphics[width=3.4in]{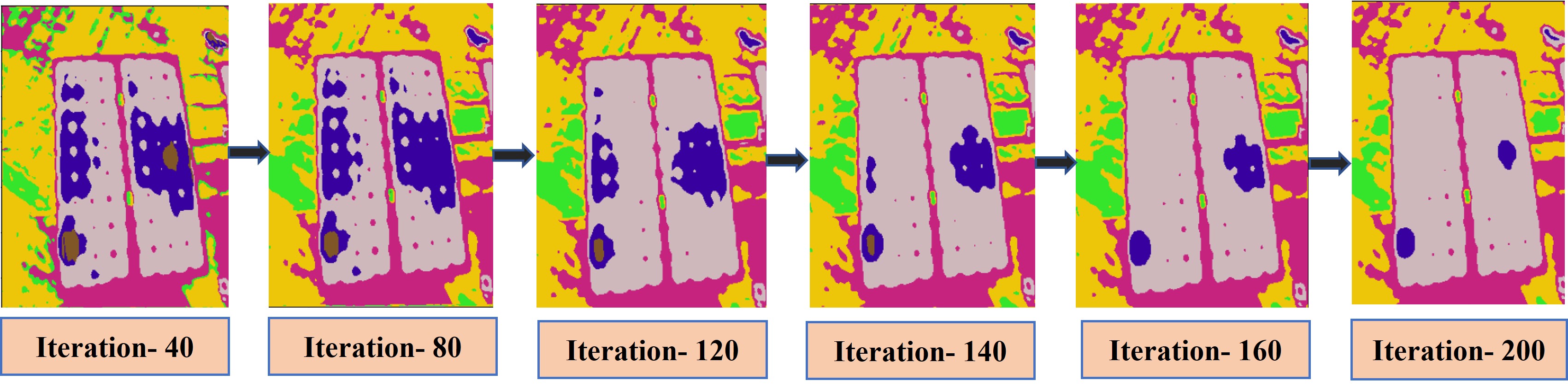}
  \caption{Implementation of backward propagation using computational loss for enhancing the cluster formation (up to 200 iterations)}
  \label{fig:epochs}
\end{figure}
\begin{figure}[h]
\centering
  \includegraphics[width=3in]{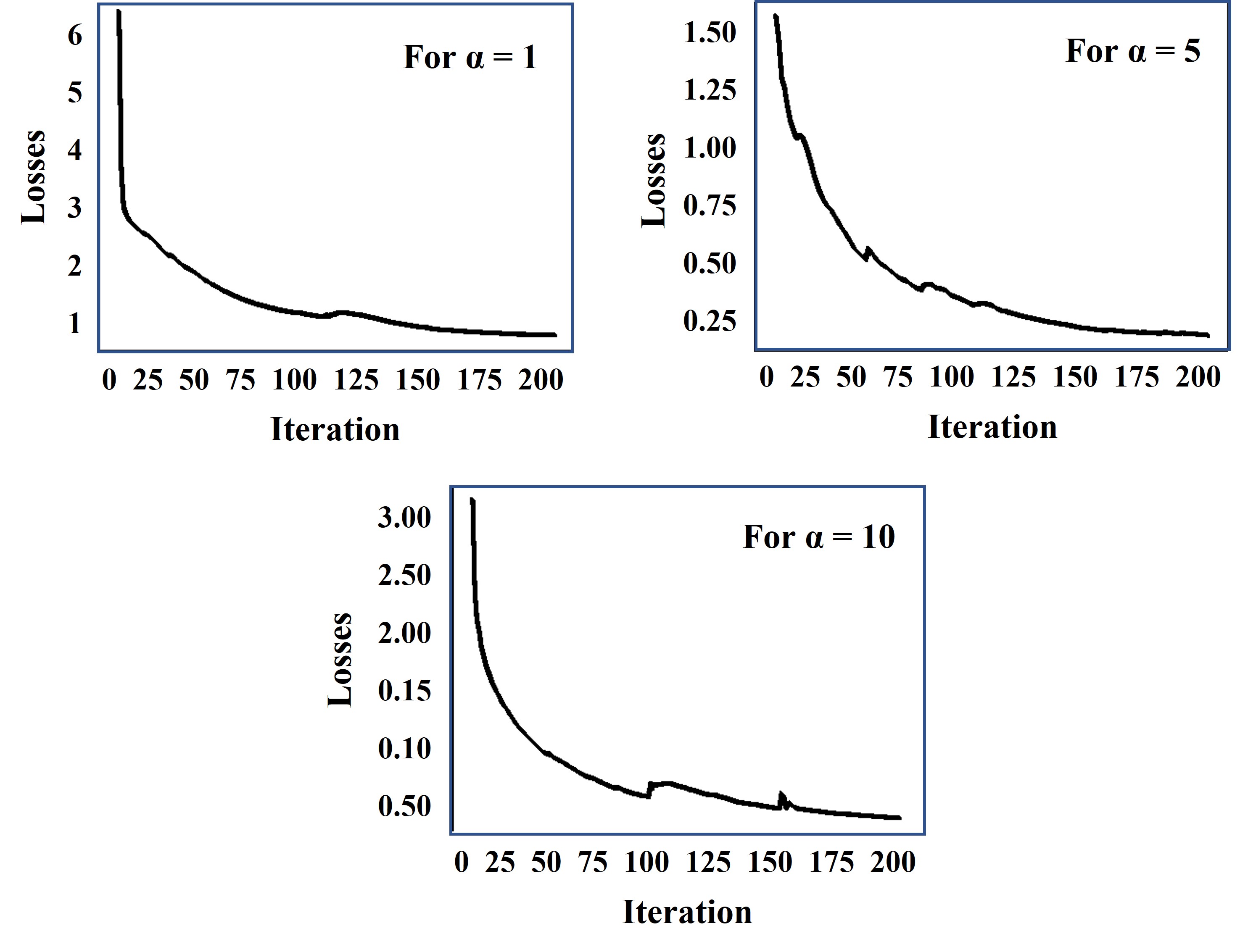}\\
  \caption{Loss v/s iteration curve for various values of $\alpha$}
  \label{fig:alpha}
\end{figure}

\begin{figure*} [h]
\centering
  \includegraphics[width=6in]{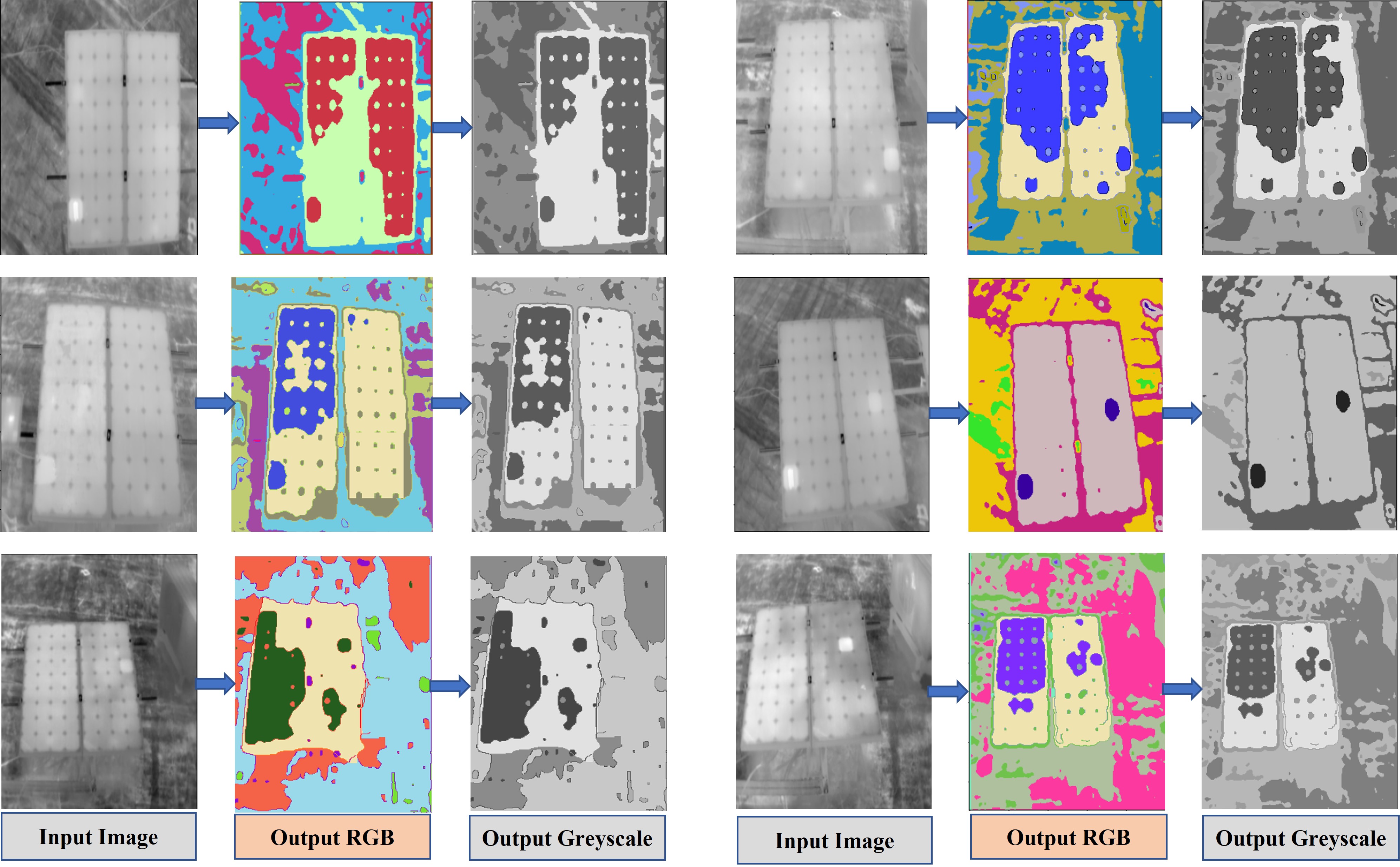}\\
  \caption{Segmentation of input greyscale thermal images of solar PV panel}
  \label{fig:Main_result}
\end{figure*}
Initially, when the greyscale thermal image of the solar PV panel is fed as input using CNN layers, the feature vector is computed and using the argmax function pseudo labels are allotted to the images as shown in \ref{fig:forward}. These initial labels are obtained using forward propagation and need further optimisation for actual cluster labels.

Thus, with the help of backward propagation the pixel labels, and their features are optimized based on iterative stochastic gradient descent. Then computed the similarity loss and spatial continuity loss to assign the same label to the pixel with similar features and spatial continuity to reduce the noises in the image segmentation process as shown in \ref{fig:epochs}.

We ran the model for 200 iterations to settle the noise in the clustered pixels and get the properly segmented image as output. In addition to that, the value for the weight balancing constant $\alpha$ in the loss function given in equation (2) needs to be fixed. Thus, we ran the model for $\alpha=1$, $\alpha=5$ and $\alpha=10$. And plotted the variation in loss function with the number of iterations as shown in Figure \ref{fig:alpha}. Here, we can see that the value of the loss function is starting with very high loss values for $\alpha=1$ and $\alpha=10$ in comparison to the $\alpha=5$. Also, at $200$ iteration, the value of the loss function is 0.508 for $\alpha=1$, 0.202 for $\alpha=5$, and 0.417 for $\alpha=5$. Thus, the loss value is comparatively low for $\alpha=5$. Thus, we fixed the $\alpha=5$ for the analysis.

After, the implementation of the model we get the final segmented solar PV panels using feature clustering. These segmented images have clustered labels differentiating the background, solar PV panel and fault defects in them. From Figure \ref{fig:Main_result}, we can see that six greyscale thermal images of solar PV panels are fed as input to the model and segmented RGB-coloured images are obtained as output. The different colours/cluster labels are allotted to objects in the image so that they are differentiable. Further, to make the analysis more reliable and implementable in the real world, we converted the segmented RGB colour image to a segmented greyscale image. This process makes it simpler to detect the faults in solar PV panels. From Figure \ref{fig:Main_result}, in segmented greyscale images, the dark spots are the faulty cells of solar panels which is easily identifiable to the human eye. Thus, we can say that the proposed framework is suitable for monitoring solar power plants. It does not require any prior training labels and ground truth labels. Also in very less computational time, the segmentation of the captured thermal images of solar PV panels is achieved and defects are identified.

\section{Conclusion}
\label{section:Conclusion}
In this paper, we proposed a novel unsupervised image segmentation algorithm based on a convolutional neural network used for segmenting faulty cells in solar photovoltaic panels. From the solar power plants, using the thermal camera infrared thermographic images of solar panels are captured. These images are pre-processed to greyscale images to extract the important features. Then the greyscale images are passed to the CNN layers which again extract out the highly important features from the input greyscale image of solar PV panels and then an argmax layer performed the differentiable task for feature clustering. The CNN layers effectively assigned the cluster labels to the pixels of the input image. Further, to achieve better clustering of similar features, backpropagation of the proposed loss function (feature similarity loss and spatial continuity loss) was applied to the normalized response of convolutional layers. As a result, it became effective in distinguishing faulty cells and normal cells in solar PV panels. To further make fault identification easy, we converted the segmented RGB image to a greyscale image, on which the dark spots are represented as the faulty cells in the panel.  Altogether, this proposed algorithm eased the segmentation of the thermal image of solar PV panels; so that defects like hot spots and snail trails can be identified. The following process and analyzed results clearly show the effectiveness of the proposed algorithm. It can easily be implemented in the real world for monitoring and maintenance of large solar power plants with the requirement of less manpower and at a low cost.

\bibliographystyle{IEEEtran}
\bibliography{main}

\end{document}